# Fast Evaporation Enabled Ultrathin Polymeric Coatings on Nanoporous Substrates for Highly Permeable Membranes

*Xiansong Shi,* [$,+] *Lei Wang,*[$,+] *Nina Yan,* [$] *Zhaogen Wang,* [$] *Leiming Guo,* [§] *Martin Steinhart,*[§,*] *and Yong Wang*[$,*]

[$] State Key Laboratory of Materials-Oriented Chemical Engineering, College of Chemical Engineering, Nanjing Tech University, Nanjing 211816, Jiangsu, P. R. China

[§] Institut für Chemie neuer Materialien, Universität Osnabrück, Barbarastr. 7, 49069 Osnabrück, Germany

[+] These authors contributed equally to this work.

ABSTRACT: Membranes derived from ultrathin polymeric films are promising to meet fast separations, but currently available approaches to produce polymer films with greatly reduced thicknesses on porous supports still faces challenges. Here, defect-free ultrathin polymer covering films (UPCFs) are realized by a facile general approach of rapid solvent evaporation. By fast evaporating dilute polymer solutions, we realize ultrathin coating (~30 nm) of porous substrates exclusively on the top surface, forming UPCFs with a block copolymer of polystyrene-*block*-poly(2-vinyl pyridine) at room temperature or a homopolymer of poly(vinyl alcohol) (PVA) at elevated temperatures. With subsequent selective swelling to the block copolymer and crosslinking to PVA, the resulting bi-layered composite structures serve as highly permeable membranes delivering ~2-10 times higher permeability in ultrafiltration and pervaporation applications than state-of-the-art separation membranes with similar rejections and selectivities. This work opens up a new, facile avenue for the controllable fabrication of ultrathin coatings on porous substrates, which shows great potentials in membrane-based separations and other areas.

The deposition of polymer solutions onto smooth solid substrates is a common route to produce polymer coatings, for example protective layers or the layers delivering specific functionalities, on a broad range of substrates. Established applications of polymeric coatings lie in the fields of painting and printing,[1] biomedicine,[2] photovoltaics,[3] optoelectronics,[4] and membrane technology.[5] The substrates utilized in the above-mentioned cases are mainly nonporous. If polymer solutions are applied to porous substrates, the nonvolatile polymeric component is commonly penetrated into the substrate pores, resulting in undesired blocking that may affect the device performance.[6-9] Thus, concentrated polymer solutions with high viscosities are typically adopted for the preparation of polymer surface films on porous supports. For example, most microfiltration and ultrafiltration membranes were produced by the deposition of concentrated polymer solutions onto macroporous nonwoven fabrics followed by precipitation in water.[10] In this way, propagation of the polymer solutions into the pores of the nonwoven fabrics is retarded, but the permeability of thus-produced membranes is correspondingly sacrificed due to huge membrane thicknesses typically exceeding 100 μm.[11] To generate polymer coatings with thicknesses in the 10 μm range on porous substrates, the pores of the latter were prefilled with a liquid (typically water), followed by deposition of a polymer solution immiscible with the liquid pre-located in the substrate pores.[12-14] However, compared to ultrathin polymer films with thicknesses down to nanometer scale, the polymer coatings with thicknesses in the 10-100 μm range exhibit great resistance to mass transfer. Therefore, ultrathin covering films with robust stability are always pursued to meet the demand of fast mass transport.[15-17] Unfortunately, directly constructing ultrathin polymer films with greatly reduced thicknesses on porous substrates still remains a huge challenge.

The formation of covering films by coating solutions on porous substrates usually involves two competing processes, that is the Brownian motion of solutes and evaporation of solvents, which determine the spatial distribution of the solute.[6] For instance, when the solvent evaporation dominates the process, the concentration gradient rises with the evaporation time, and the solute accumulates at the solution/air interface to form a thin coverage in the absence of infiltration. Undoubtedly, the final structure and solute distribution greatly depend on the evaporation rate of solvent.[18] Therefore, we speculate that the control over the evaporation rate of solvent, which can be easily realized in the experiment, may open up a new avenue to effectively govern the polymer distributions and film thicknesses, thus offering the possibility to engineer ultrathin polymer films with significantly decreased thicknesses on porous substrates.

Here we report the preparation of ultrathin polymeric covering films (UPCFs) enabled by fast evaporation of solvent and demonstrate their excellence as platforms for producing highly permeable membranes. Ultrafast evaporation of solvents induces the generation of a thin skin layer on anodic aluminum oxide (AAO) substrates, preventing the infiltration of solutes into substrate pores. With the subsequent treatments to as-prepared UPCFs, such as the cavitation by selective swelling of block copolymers (BCPs) and crosslinking of poly(vinyl alcohol) (PVA), the resultant BCP and PVA membranes show excellent ultrafiltration and pervaporation performance characterized by high permeability at no or little expense of selectivity. Our findings establish a new platform to develop ultrathin covering films with diverse building blocks for ultrafast membrane separations.

The commercial AAO substrate utilized in this work consists of a thin size-selective layer with 100-nm thickness containing pores with a diameter of ~20 nm (thereafter referred to as "size-selective AAO layer", Figure S1). The size-selective AAO layer is supported by a 60-µm thick AAO layer containing aligned cylindrical pores with diameters scattering about 200 nm. To prepare uniform and continuous UPCFs having thicknesses down to ~30 nm, we deposited dilute solutions of gelating polymers onto the size-selective AAO layers under conditions of fast solvent evaporation (Figure 1a). Thus, a bulk reservoir of the deposited solution covers the size-selective AAO layer from which the solvent evaporates. Diffusion processes in the solution are initially fast enough to prevent the occurrence of pronounced concentration gradients; the evaporation kinetics is determined by evaporative flux. Thereafter, mass transport slows down due to the progressing decrease in the solvent concentration. Solvent depletion at the evaporation surface is then no longer compensated by diffusion and a skin of the nonvolatile solutes forms at the solution surface.[19-20] Here, the formation of a polymeric skin at the surface of the solution is also promoted by the evaporation-induced temperature drop and by the tendency of the selected polymers to gelate. After complete solvent evaporation, the polymeric skin is thus converted into a defect-free and continuous UPCF. The deposited polymer is mainly located in the UPCF while deposition of the polymeric component into AAO pores is largely prevented.

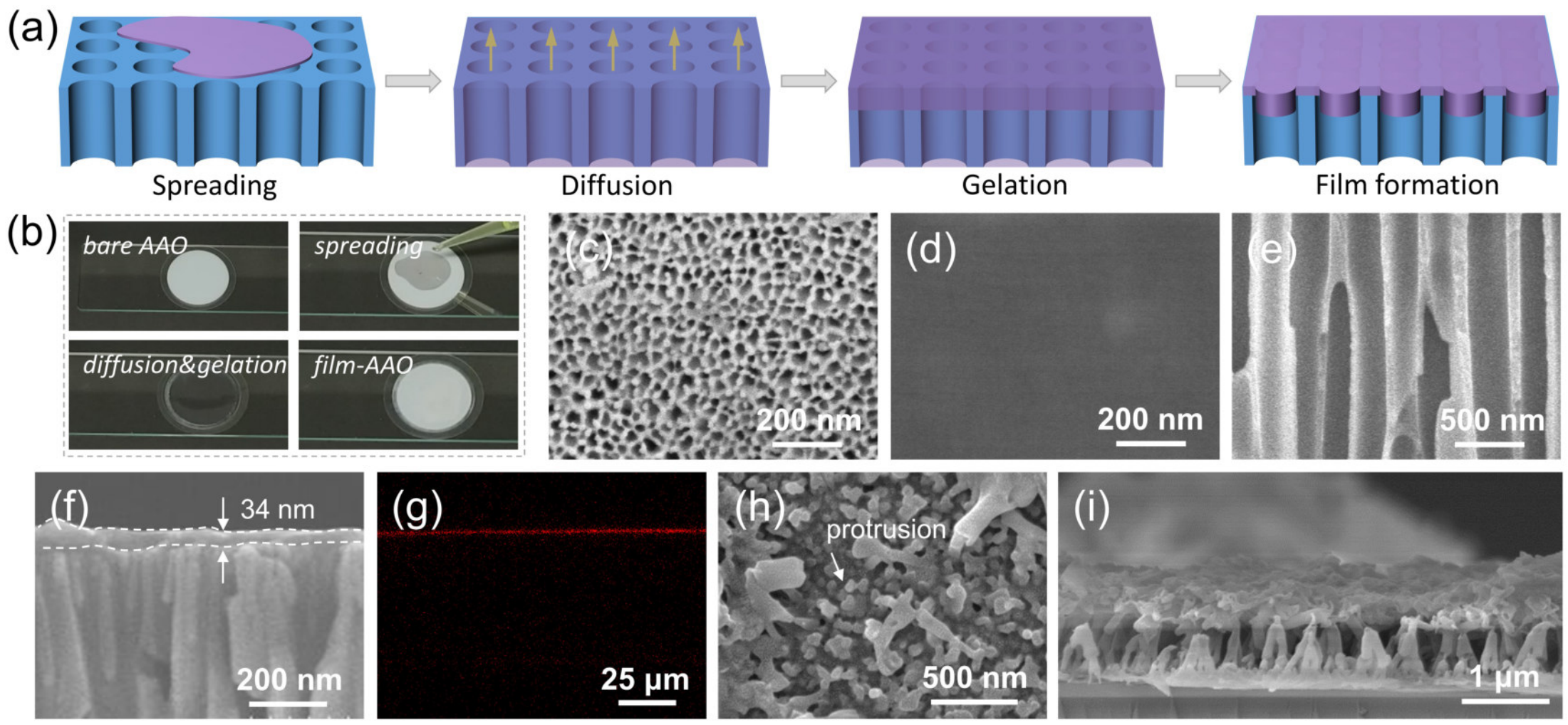

**Figure 1.** Fabrication of PS-*b*-P2VP UPCFs. (a) Schematic diagram for the film preparation. (b) Digital images during the film preparation. Surface SEM images of the bare AAO substrate (c) and UPCF (d). (e, f) Cross-sectional SEM images of the UPCF. (g) Cross-sectional fluorescent image. Bottom (h) and cross-sectional SEM images of the UPCF after removing AAO substrates. The UPCF was prepared by coating a 0.05 wt% PS-*b*-P2VP/$CS_2$ solution.

We prepared UPCFs from the block copolymer (BCP) of Polystyrene-*block*-poly(2-vinyl pyridine) (PS-*b*-P2VP), which can easily be cavitated by selective swelling-induced pore generation so that mesoporous PS-*b*-P2VP-UPCFs (*m*UPCFs) are accessible.[21-24] Here, carbon disulfide ($CS_2$) was deliberately selected as the solvent due to its high volatility (vapor pressure 309 mmHg at 21°C) and its low boiling point of 46.5°C.[25] Solutions of atactic PS in $CS_2$ are known to form unusually stable gels. Since atactic PS cannot crystallize, gelation is related to specific solvent-polymer interactions.[26-27] The PS is preferentially exposed to $CS_2$ as PS possesses a higher solubility in this solvent than P2VP.[28] 100 μL portions of PS-*b*-P2VP/$CS_2$ solutions were spread on the size-selective AAO layers at 20±5°C (Figure 1b). The initially opaque AAO substrates immediately turned transparent, indicating fast imbibition. For a concentration of 0.1 wt% PS-*b*-P2VP, the AAO turned opaque again after ~70 s, implying complete evaporation of $CS_2$ (Figure 1b, film-AAO). After coating a 0.05 wt% PS-*b*-P2VP/$CS_2$ solution, the obtained UPCF shows a smooth and continuous surface morphology, and no cracks, pinholes, or other defects are visible (Figure 1d). The dense structure contrasts sharply with the bare AAO substrate (Figure 1c). Moreover, the coating solution with a concentration of 0.03 wt% and 0.1 wt% is also applicable to form a defect-free UPCF (Figure S2). With the fast evaporation of $CS_2$, the PS-*b*-P2VP is predominantly located in the UPCF, evidenced by the unimpeded pores without any blocking (Figure 1e, Figure S3). The formed UPCFs exhibit tunable thicknesses in the range of ~30-215 nm, with the solution concentration rising from 0.03 to 0.1 wt% (Figure 1f, Figure S4). The cross-sectional fluorescent image also demonstrates the formation of an ultrathin PS-b-P2VP film (Figure 1g). AAO substrates were then degraded to clearly observe the film structure, as shown in Figure 1h, i. Short protrusions with a length of ~650 nm exist on the bottom of PS-*b*-P2VP-UPCFs, which we interpret as negative replicas of the pore mouths of size-selective AAO layer.

Assuming that the porosity of AAO substrates amounts to 50%, the substrate correspondingly has an integrated pore volume of ~15 μL. Thus, 100 μL PS-*b*-P2VP/$CS_2$ solution ensures a complete filling of the AAO pores, and the excess solution forms a bulk reservoir covering the size-selective AAO layer. This speculation is also demonstrated by the appearance change of AAO during preparation as discussed above. In this case, PS-*b*-P2VP should uniformly locate on the surface and entire pore walls of AAO substrate after $CS_2$ evaporation. Surprisingly, we can only observe a PS-*b*-P2VP film on the top layer of AAO substrates. The UPCF formation mechanism is given as follows. After solution spreading, the $CS_2$ evaporation will not only result in the depletion of the solvent and the enrichment of PS-*b*-P2VP at the solution surface but also give a temperature drop. Rapid gelation of the PS blocks of PS-*b*-P2VP in the presence of $CS_2$, possibly superimposed by vitrification of PS and/or P2VP, will transform the solution from viscoelastic fluid to viscoelastic solid, generating a thin PS-*b*-P2VP skin that separates PS-*b*-P2VP/$CS_2$ solution from air.[6] Benefiting from the excellent compatibility between PS and $CS_2$, the succeeding evaporation of $CS_2$ is unobstructed with the existence of a PS-*b*-P2VP skin. Complete evaporation of the $CS_2$ transforms the thin PS-*b*-P2VP skin into an UPCF on the AAO substrate. Besides, during the infiltration of solution into AAO pores, irreversible adsorption of PS-*b*-P2VP molecules on the size-selective AAO layer may hinder further PS-*b*-P2VP molecules to enter the narrow AAO pore necks according to the findings reported by Karagiovanaki *et al.*[29] Additionally, the fast evaporation of $CS_2$ rapidly rises the solution concentration, contributing to the immobilization of PS-*b*-P2VP on the top of AAO. These results synergistically lead to the generation of continuous films having short protrusions underneath them. The wetting of glass slides, which was used to hold AAO substrates, by $CS_2$ solution but no PS-*b*-P2VP on them evidences the proposed mechanism.

To convert nonporous PS-*b*-P2VP-UPCFs into *m*UPCFs, we introduce a nondestructive pore-making strategy, that is selective swelling (Figure 2a). Specifically, a PS-*b*-P2VP-UPCF prepared by deposition of a solution of 0.05 wt% PS-*b*-P2VP in $CS_2$ was treated in ethanol at 70°C for 1 h. Notably, *m*UPCFs swollen at different temperatures exhibit a similar structure, showing a continuous-spongy network of curved, interconnected PS-*b*-P2VP cylinders with a diameter of ~40 nm (Figure 2b, Figure S5). Inspection of SEM images reveals that the *m*UPCFs contain continuous mesopore systems with diameters ranging from ~20 to ~60 nm. Though the size of these pores is larger than that of the size-selective AAO layer, the stagger stacking of porous PS-*b*-P2VP layers and AAO top layers at the interface gives a significantly reduced pore size, as shown in Figure 2c, d and Figure S6. Thus shrunken channels will promote the selectivity of the PS-*b*-P2VP-*m*UPCF/AAO composite membranes, and similar improvements in selectivity induced by the stagger stacking also can be found in other nanoporous materials, such as covalent organic frameworks (COFs) stacked in the offset eclipsed fashion.[30-31] More importantly, the mesopores on the both sides of the interface deliver a low resistance to solvent and solute, thus promising an enhanced selectivity without noticeably sacrificing permeability.

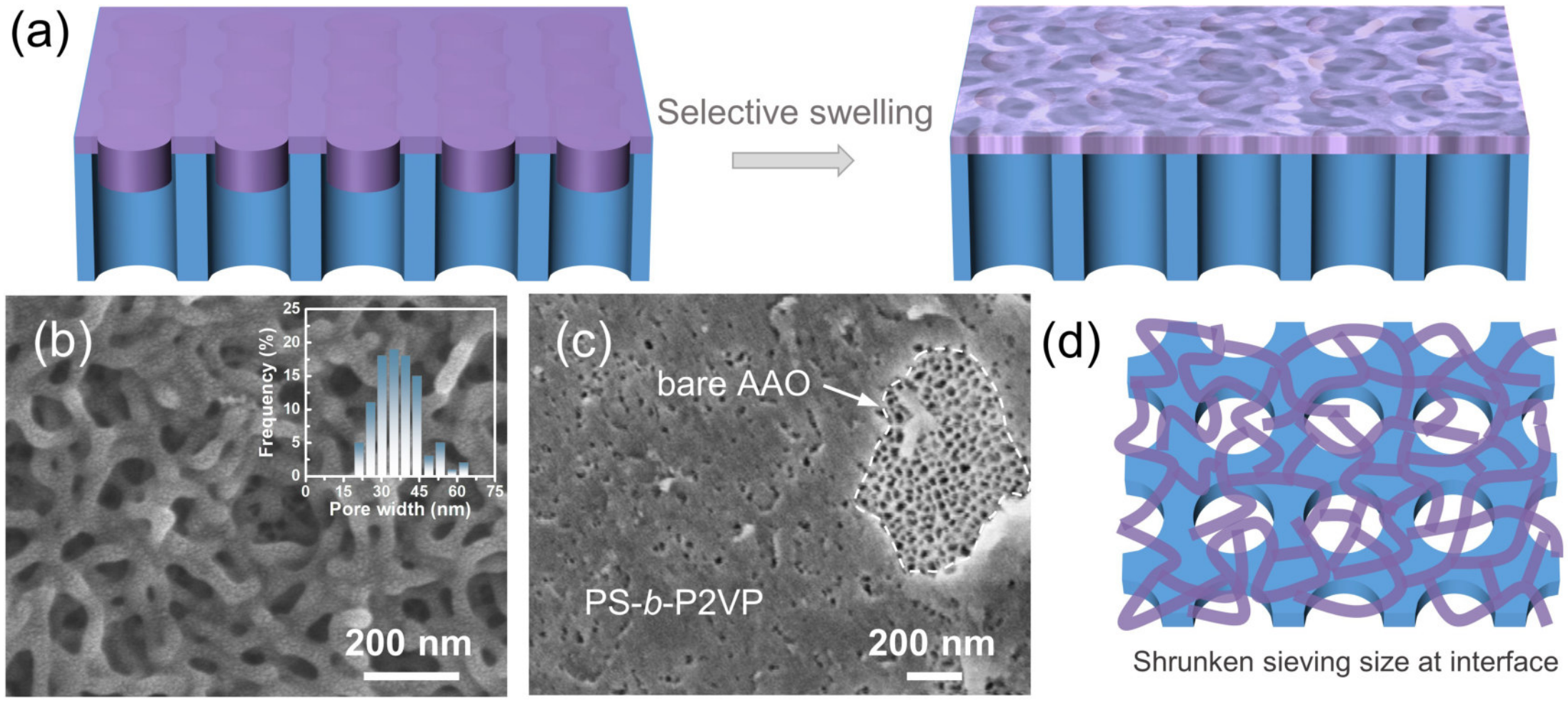

**Figure 2.** Selective swelling of PS-*b*-P2VP-UPCF/AAO composite membranes. (a) Schematic diagram of the selective swelling. Surface SEM images of the pristine (b) and polished (c) PS-*b*-P2VP-*m*UPCF/AAO membranes. The membrane was prepared with a 0.05 wt% precursor solution followed by swelling in 70°C ethanol for 1 h, and the inset in (b) shows the pore size distribution. (d) Illustration of the shrunken sieving sizes induced by stagger stacking.

As shown in Figure 3a, the separation performance of PS-*b*-P2VP-*m*UPCF/AAO composite membranes is relatively robust against variations of the preparation conditions, including PS-*b*-P2VP concentrations as well as swelling temperatures and durations. For instance, rising the PS-*b*-P2VP concentration in the initially applied PS-*b*-P2VP/$CS_2$ solutions from 0.03 to 0.1 wt% results in a moderate decrease in water permeance from ~1275 to ~1042 L $m^{-2}$ $h^{-1}$ $bar^{-1}$. Increasing the temperature during selective swelling leads to more pronounced cavitation of the *m*UPCFs and in turn to increased water permeance.[32] When the swelling temperature rises from 65 to 75°C, the water permeability increases from ~915 to ~1099 L $m^{-2}$ $h^{-1}$ $bar^{-1}$. Moreover, the swelling duration causes negligible influence on the water permeance. In all these cases, the rejections to bovine serum albumin (BSA, $M_w$=67 kDa) are >92% and remain nearly unchanged. It is worth noting that *m*UPCF/AAO composite membranes show tight rejections to proteins while they still exhibit a high water permeance of up to almost half the permeance of bare AAO substrates, which show no rejection to BSA. The *m*UPCF/AAO composite membrane also displays a rejection of ~95% and ~22% to ovalbumin (OVA, $M_w$=45 kDa) and cytochrome *C* (Cyt. *C*, $M_w$=12.6 kDa), respectively, giving an approximatively molecular-weight-cut-off (MWCO) of ~42 kDa (Figure 3b). Apart from abovementioned high porosities and stagger structures, this excellent separation performance may originate from polar and water-permeable P2VP blocks as well.[33] Further, the benefit of $CS_2$ adoption can be easily perceived by comparing the separation performance of membranes prepared with other solvents (Figure S7). Given the tight selectivity of our membranes, the resultant *m*UPCF/AAO composite membranes are applicable to recover valuable nanoparticles from water. As a demonstration, we implemented the concentration of CdTe quantum dots (QDs) with a diameter of 4 nm dispersed in water, as shown in Figure 3c. The emission spectrum of the QDs in feed shows a strong peak at ~570 nm and intense fluorescence signal (inset in Figure 3c). In contrast, the filtrate exhibits no peak and fluorescence signal, indicating a prominent repulsion of QDs by our membranes. The increased fluorescence of retentate reveals that the size-sieving rather than adsorption plays a dominating role in the QDs recovery. Compared to our previously reported BCP-based membranes[23, 34-35] and other membranes prepared by various materials,[36-44] the PS-*b*-P2VP-*m*UPCF/AAO composite membranes show outstanding selectivities with ~2-10 times higher water permeance (Figure 3d), indicating their high efficiency for membrane-based separations, such as the protein purification, nanoparticle concentration, and recovery of nano-sized precious metals from water.

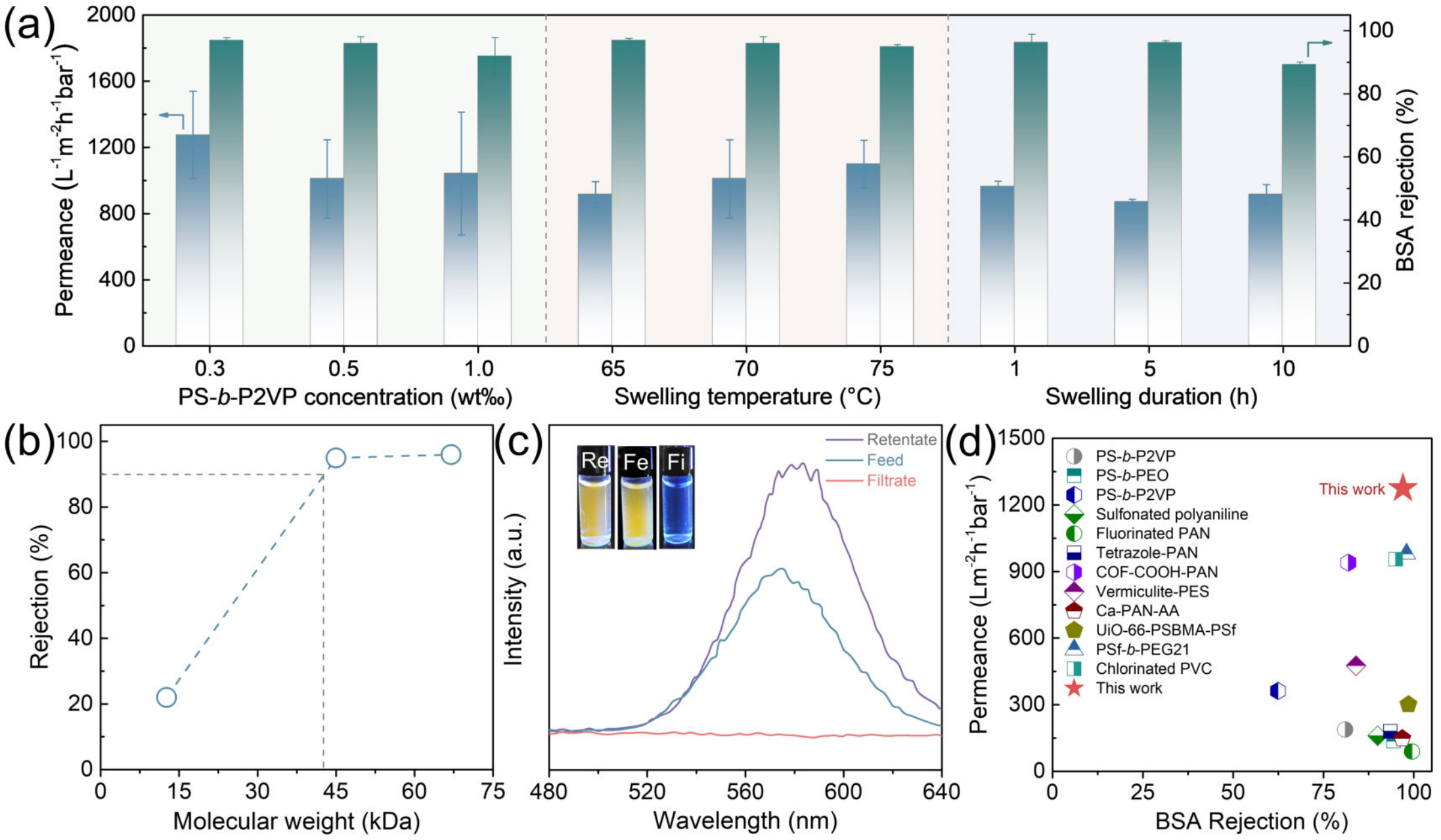

**Figure 3.** Separation performance of PS-*b*-P2VP-*m*UPCF/AAO composite membranes. (a) Performances with various PS-*b*-P2VP concentrations, swelling temperatures, and durations. (b) Rejections to various objects with different molecular weight. (c) Emission spectra of the feed, filtrate and retentate obtained by concentrating CdTe QDs. (d) Comparison on the separation performance of various membranes. Inset in (c) shows fluorescent photographs of the feed (Fe),

filtrate (Fi), and retentate (Re). The results shown in panels (b), (c), and (d) were obtained from the membrane prepared with a 0.05 wt% precursor solution followed by swelling in 70°C ethanol for 1 h.

Post-cavitation of dense PS-*b*-P2VP-UPCFs leads to mesoporous UPCFs that can be adopted for highly permeable ultrafiltration. Alternatively, as-coated dense UPCFs are considered as a promising candidate for pervaporation through the solution-diffusion model.[45-46] To this end, we prepared cross-linked poly(vinyl alcohol) films (PVA-UPCFs) for the recovery of ethanol from dilute aqueous solution. The rapid evaporation of solvent results in uniform and continuous PVA-UPCFs supported by AAO substrates (Figure 4a, b). The control over solution volume during coating enables a tunable film thickness. Concretely, we can obtain films with a thickness of ~307 and ~585 nm when applying a solution volume of 100, and 200 μL, respectively. After cross-linking the PVA with maleic anhydride,[47] the dehydration performance of the obtained PVA-UPCF/AAO composite membranes were investigated. Taking advantage of the fact that water diffuses faster through PVA than ethanol,[48] we used PVA-UPCF/AAO composite membranes to concentrate water at the permeate side, as illustrated in Figure 4c. As a result, while the water content in the feed amounts to ~10 wt%, water is successfully concentrated to >75 wt% in the permeate after pervaporation through our membranes. The separation factor and the flux are determined to be 14.7 and 600 g $m^{-2}$ $h^{-1}$ for the PVA-UPCF/AAO composite membrane resulted from 100 μL PVA solution (Figure 4d). Increasing the volume of the deposited PVA solution to 200 μL improves the separation factor to 27.6 while a high flux of 463 g $m^{-2}$ $h^{-1}$ is still retained. In terms of permeability such dehydration performances are excellent compared to previously reported pervaporation membranes prepared by other methods.[49-51]

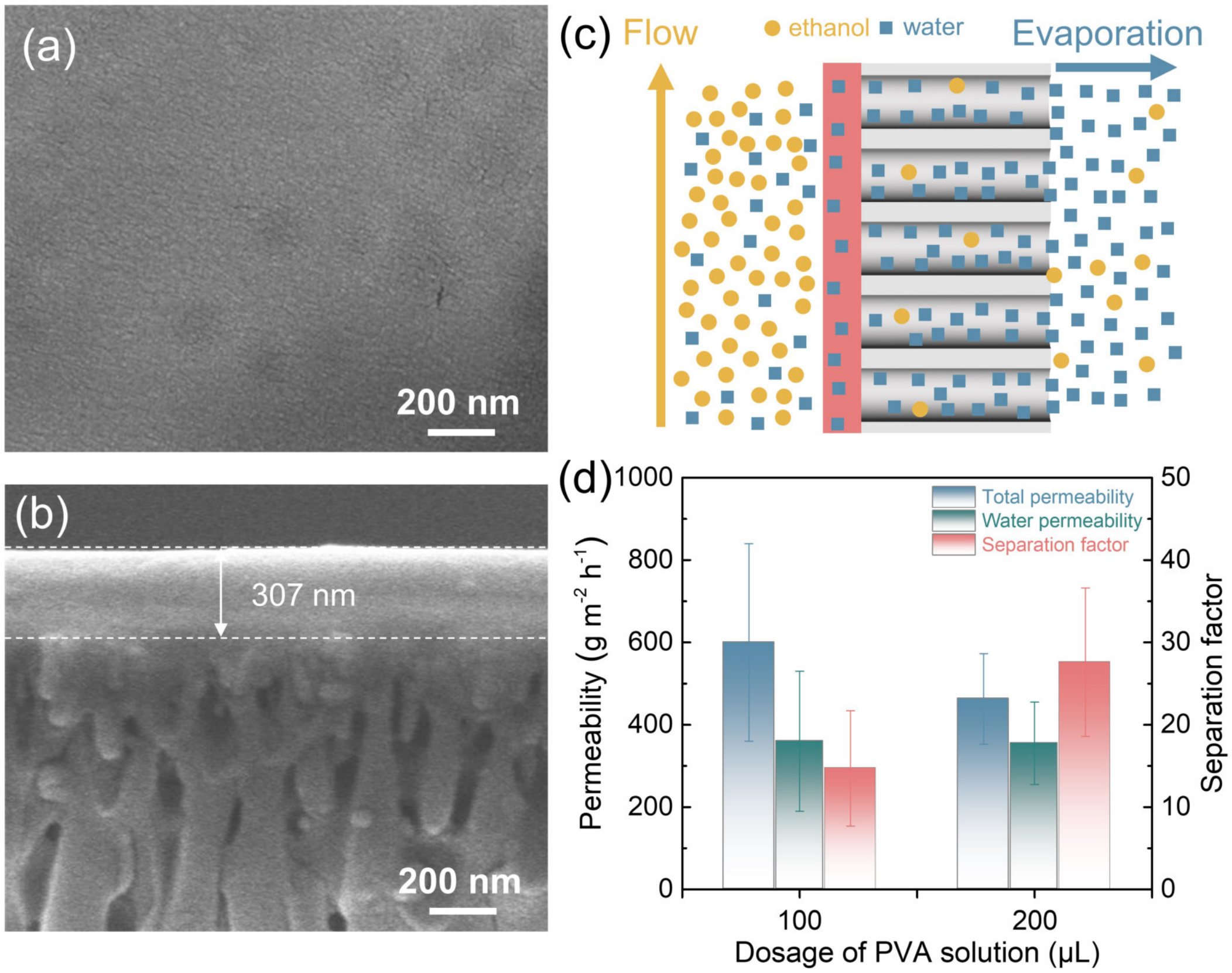

**Figure 4.** Cross-linked PVA-UPCF/AAO composite membranes for the dehydration of ethanol by pervaporation. Surface (a) and cross-sectional (b) SEM images of the membrane prepared with100 μL PVA solution. (c) Schematic diagram for the pervaporation. (d) Permeabilities and separation factors of membranes prepared with various dosages of PVA solution.

In summary, rapidly evaporating dilute polymeric solution composed of solvent with a high volatility is a promising strategy to produce ultrathin films that are capable of acting as highly permeable separation membranes. The formation of thin films on the top of porous substrates observed in this work can be mostly ascribed to the fast gelation of polymeric precursors, which induces the generation of a skin layer that prevents the further polymer infiltration into substrate pores. The preparation methodology proposed here exhibit an excellent universality, evidenced by the fabrication of PS-*b*-P2VP and PVA thin films. The selective swelling and crosslinking allow the PS-*b*-P2VP and PVA films to become separation membranes for effectively purifying proteins and recovering valuable nanoparticle by ultrafiltration as well as concentrating ethanol from its aqueous solution by pervaporation, respectively. Our results provide a new avenue to directly construct ultrathin polymeric films on porous substrates from their dilute precursor solutions, and the suggested strategy is generally applicable to prepare diverse polymeric films for extensive applications in many fields.

ASSOCIATED CONTENT

**Supporting Information**

The following files are available free of charge.

Scheme of the setup used for pervaporation experiments; structure of AAO substrates; surface and cross-sectional SEM images of PS-*b*-P2VP-UPCFs/AAO composite membranes; surface SEM images of PS-*b*-P2VP-*m*UPCFs/AAO composite membranes; separation performance of PS-*b*-P2VP-*m*UPCFs/AAO composite membranes prepared with various solvents. (PDF)

Video of the PS-b-P2VP-UPCFs/AAO composite membrane preparation. (mp4)

AUTHOR INFORMATION

**Corresponding Author**

**Martin Steinhart**- Institut für Chemie neuer Materialien, Universität Osnabrück, Barbarastr. 7, 49069 Osnabrück, Germany; E-mail: martin.steinhart@uni-osnabrueck.de

**Yong Wang-**State Key Laboratory of Materials-Oriented Chemical Engineering, College of Chemical Engineering, Nanjing Tech University, Nanjing 211816, Jiangsu, P. R. China; E-mail: yongwang@njtech.edu.cn

**Author**

**Xiansong Shi**-State Key Laboratory of Materials-Oriented Chemical Engineering, College of Chemical Engineering, Nanjing Tech University, Nanjing 211816, Jiangsu, P. R. China

**Lei Wang**-State Key Laboratory of Materials-Oriented Chemical Engineering, College of Chemical Engineering, Nanjing Tech University, Nanjing 211816, Jiangsu, P. R. China

**Nina Yan**-State Key Laboratory of Materials-Oriented Chemical Engineering, College of Chemical Engineering, Nanjing Tech University, Nanjing 211816, Jiangsu, P. R. China

**Zhaogen Wang**-State Key Laboratory of Materials-Oriented Chemical Engineering, College of Chemical Engineering, Nanjing Tech University, Nanjing 211816, Jiangsu, P. R. China

**Leiming Guo**- Institut für Chemie neuer Materialien, Universität Osnabrück, Barbarastr. 7, 49069 Osnabrück, Germany

**Author Contributions**

The manuscript was written through contributions of all authors. All authors have given approval to the final version of the manuscript. [+]X. S. and L. W. contributed equally to this work.

**Notes**

The authors declare no competing financial interest.

ACKNOWLEDGMENT

Financial support from the Natural Science Fund for Distinguished Young Scholars (21852503) is gratefully acknowledged. We also thank Program of Excellent Innovation Teams of Jiangsu Higher Education Institutions and the Project of Priority Academic Program Development of Jiangsu Higher Education Institutions (PAPD) for support. M. S. thanks the European Research Council (ERC-CoG-2014, project 646742 INCANA) and the German Research Foundation (INST 190/164-1 FUGG) for funding.

## Graphic abstract

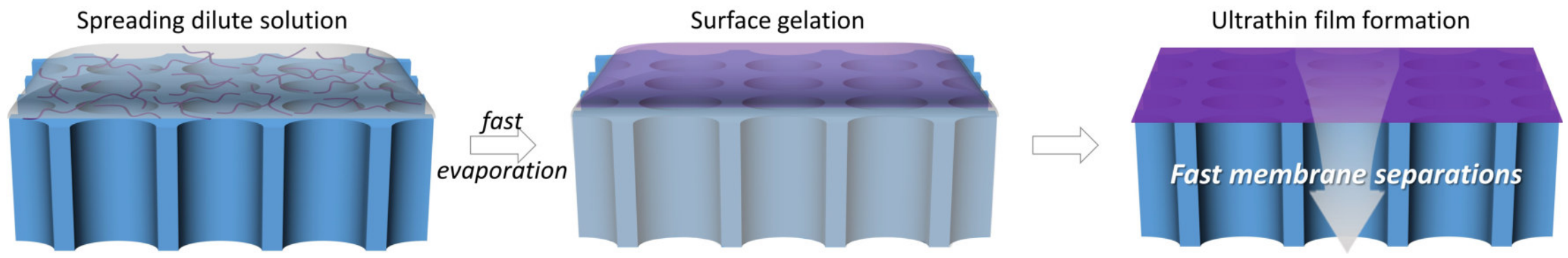

Supporting Information

# Fast Evaporation Enabled Ultrathin Polymeric Coatings on Nanoporous Substrates for Highly Permeable Membranes

*Xiansong Shi,* [$,+] *Lei Wang,*[$,+] *Nina Yan,* [$] *Zhaogen Wang,* [$] *Leiming Guo,* [§]

*Martin Steinhart,*[§,*] *and Yong Wang*[$,*]

[$] State Key Laboratory of Materials-Oriented Chemical Engineering, and College of Chemical Engineering, Nanjing Tech University, Nanjing 211816, Jiangsu, P. R. China

[§] Institut für Chemie neuer Materialien, Universität Osnabrück, Barbarastraße 7, 49069 Osnabrück, Germany

[+] These authors contributed equally to this work.

## Experimental section

**Materials.** Polystyrene-*block*-poly(2-vinyl pyridine) (PS-*b*-P2VP, $M_n$(PS)=53 kg mol$^{-1}$, $M_n$(P2VP)=21 kg mol$^{-1}$; polydispersity index: 1.17) was supplied by Tubang Polymer Materials Co., Ltd and washed in ethanol at 50°C for one week to remove any soluble components and then dried at 50°C for two days. Poly(vinyl alcohol) (PVA) (degree of polymerization: 1750±50) was obtained from Sinopharm Chemical Reagent Co., Ltd., China. Maleic anhydride (MAH) and sulfuric acid of analytical grade provided by Shanghai Lingfeng Chemical Solvent Factory were used as received. As anodic aluminum oxide (AAO) substrates, we used Whatman Anodisc inorganic filter membranes with a diameter of 25 mm and a nominal pore size of 20 nm. Bovine serum albumin (BSA, $M_w$=67 kDa) with a purity >97% as well as ovalbumin (OVA, $M_w$=45 kDa) with a purity >90% were both purchased from Aladdin and used without further purification. Cytochrome *C* (Cyt. *C*, $M_w$=12.4 kDa) was supplied by Sigma Aldrich; CdTe quantum dots (QDs) (diameter: ~4 nm; concentration: 10 g L$^{-1}$) aqueous solution was supplied by Janus New-Materials Co., Ltd. Deionized (DI) water was used in all tests. Phosphate buffer saline (PBS) solutions were prepared by dissolving the PBS tablets (MP Biomedicals, LLC) in DI water. Potassium hydroxide (KOH), carbon disulfide ($CS_2$), chloroform, and toluene were obtained from local suppliers and used as received.

**Preparation of PS-*b*-P2VP-UPCF/AAO composite membranes.** The applied solutions of PS-*b*-P2VP in $CS_2$ were filtered through polytetrafluoroethylene (PTFE) filters with a pore diameter of 220 nm for three times. The concentration of PS-*b*-P2VP solution was typically 0.05 wt% if not otherwise stated. The AAO substrates were placed on glass slides in such a way that the size-selective AAO layer pointed upwards. Then, 100 μL of the PS-*b*-P2VP/$CS_2$ solutions were quickly dropped onto the size-selective AAO layer at a temperature of 20±5°C and a relative humidity of 40±10%. After sufficient evaporation of $CS_2$, the obtained composite membranes were stored at room temperature before further use.

**Preparation of PS-*b*-P2VP-*m*UPCF/AAO composite membranes.** The dense PS-*b*-P2VP-UPCF/AAO composite membranes were immersed into warm ethanol for various durations followed by air drying, thus producing mesoporous PS-*b*-P2VP-UPCF (PS-*b*-P2VP-*m*UPCF) /AAO composite membranes. We should note that the swelling agent of ethanol is recyclable.

**Preparation of cross-linked PVA-UPCF/AAO composite membranes.** 0.03 g PVA was dissolved in 7 g of DI water at 85°C. The aqueous solution was filtered through a PTFE filter. Then, a portion of 2 g of the aqueous PVA solution was mixed with 0.86 g *n*-propyl alcohol at 85°C, yielding a solution with a PVA concentration of 0.3 wt%. The AAO substrates were placed on glass slides in such a way that the size-selective AAO layer pointed upwards and then heated to 85°C prior to the deposition of the PVA solutions. During cross-linking of PVA, the pH value of an aqueous solution of 0.2 wt% MAH was adjusted to ~1 using sulfuric acid. The PVA-UPCF/AAO composite membranes were immersed into the MAH solution at 35°C for 1 h to crosslink the PVA, thoroughly washed with deionized water, dried at 55°C for at least 1 h followed by drying at 120°C for 30 min.

**Characterization.** Scanning electron microscope (SEM) images were acquired with a field-emission SEM Hitachi S-4800 using an accelerating voltage <5 kV. Prior to SEM imaging the samples were sputter-coated with a thin layer of platinum. Cross-sectional samples were prepared by fracturing in liquid nitrogen. To observe the formed PS-*b*-P2VP-UPCF, the composite membranes were also soaked in 40 wt% KOH aqueous solution followed by sufficient water washing to remove AAO supports. A Leica TCS/SP2 confocal microscope was also applied to analyze the distribution of PS-*b*-P2VP-UPCF in the produced composite membrane.

**Filtration tests.** Water permeance and retentions of BSA, OVA, and Cyt. *C* were determined for PS-*b*-P2VP-UPCF/AAO composite membranes as well as for PS-*b*-P2VP-*m*UPCFs/AAO composite membranes at room temperature in a stirred cell module (Amicon 8010, Millipore) under a pressure of 0.4 bar. To ensure a stable permeability, a 25-minute pre-pressing to the membrane samples was employed at a pressure of 0.5 bar prior to the filtration tests. For the retention tests, feed solutions of BSA, OVA and Cyt. C were prepared by dissolving them in PBS aqueous solutions at concentrations of 0.5 g $L^{-1}$, 0.5 g $L^{-1}$, and 0.04 g $L^{-1}$, respectively. The retention rates were determined by measuring the concentrations of the feed as well as the filtrate with a UV-vis absorption spectrophotometer (NanoDrop 2000c, Thermo). Photoluminescence spectra of the feed, filtrate, and retentate of CdTe QDs solutions were measured using a Varian Cary Eclipse spectrophotometer operating with an excitation wavelength of 450 nm and a voltage of 700 V.

**Pervaporation tests.** A scheme of the pervaporation setup is shown in Scheme S1. Dehydration of water/ethanol mixtures by cross-linked PVA-UPCF composite membranes was tested with a water content of 10 wt% at 75°C. The feed liquid with a constant flow rate of 500 mL min$^{-1}$ was driven through the membrane at atmospheric pressure by a peristaltic pump and flowed back to the feed tank. The pressure of the permeate side was maintained below 200 Pa by a vacuum pump. The permeate was condensed in a cooling trap cooled with liquid nitrogen. The compositions of the feed and permeate were analyzed by a gas chromatograph (GC-2014, Shimadzu Corporation) equipped with a thermal conductivity detector (TCD) and a packed column of Parapak-Q. The pervaporation performance was evaluated by the flux ($J$, g·m$^{-2}$·h$^{-1}$) and the separation factor ($\alpha$), which can be defined as follows:

$$J=m/(At) \qquad (1)$$

$$\alpha=(y_W x_E)/(y_E x_W) \qquad (2)$$

where $m$ is mass of the permeate, $t$ is the time interval, $A$ is the effective membrane area, which is 7.5×10$^{-5}$ m$^2$ in our experiment, $y_W$, $y_E$, $x_E$, $x_W$ are the mass fractions of water and ethanol in permeate and feed, respectively.

Figures

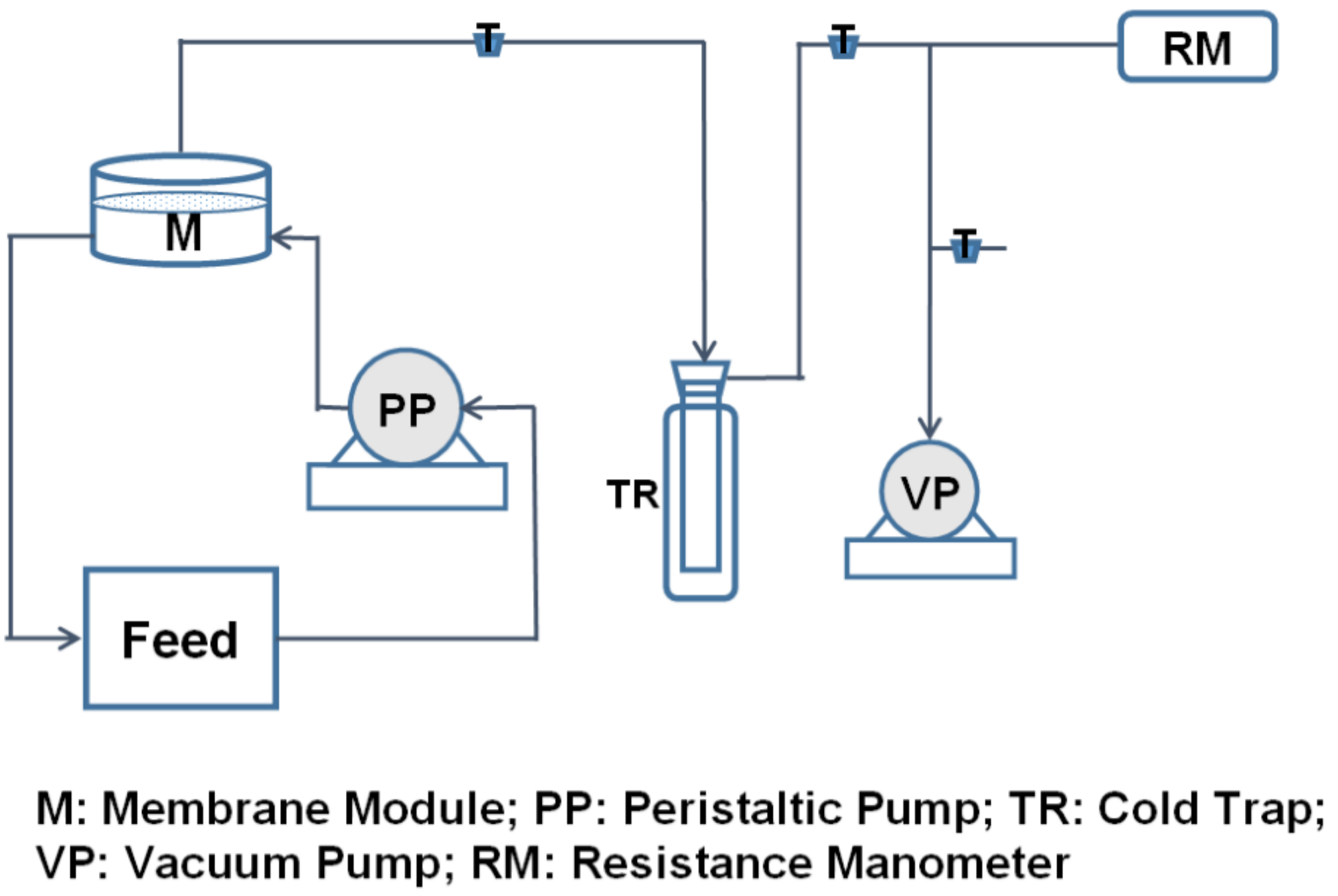

**Scheme S1.** Scheme of the setup used for pervaporation experiments.

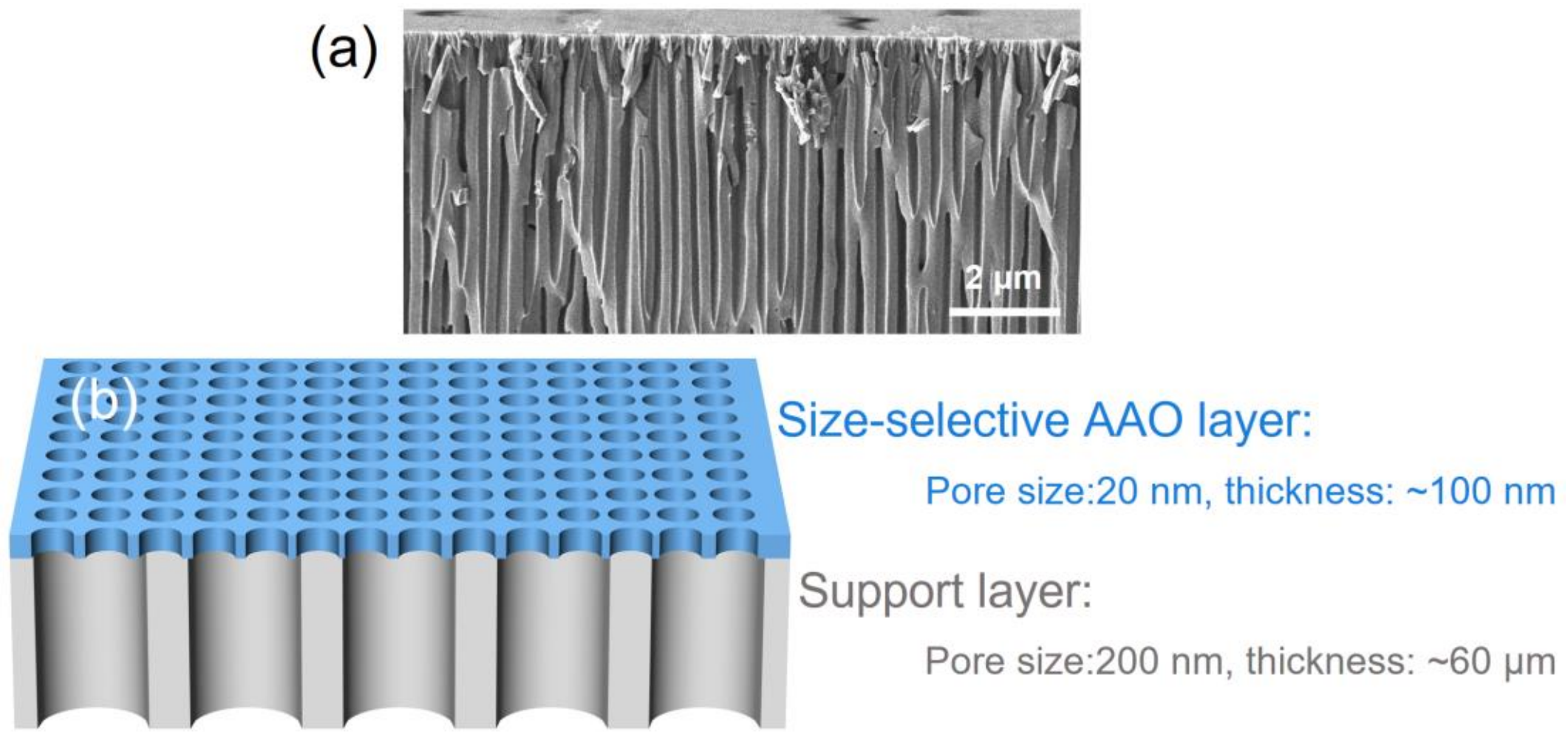

**Figure S1.** Cross-sectional SEM image (a) and illustration (b) of an as-received AAO membrane showing the thin size-selective AAO layer (pore diameter ~20 nm) at the top and the underlying AAO membrane with pore diameters scattering about 200 nm.

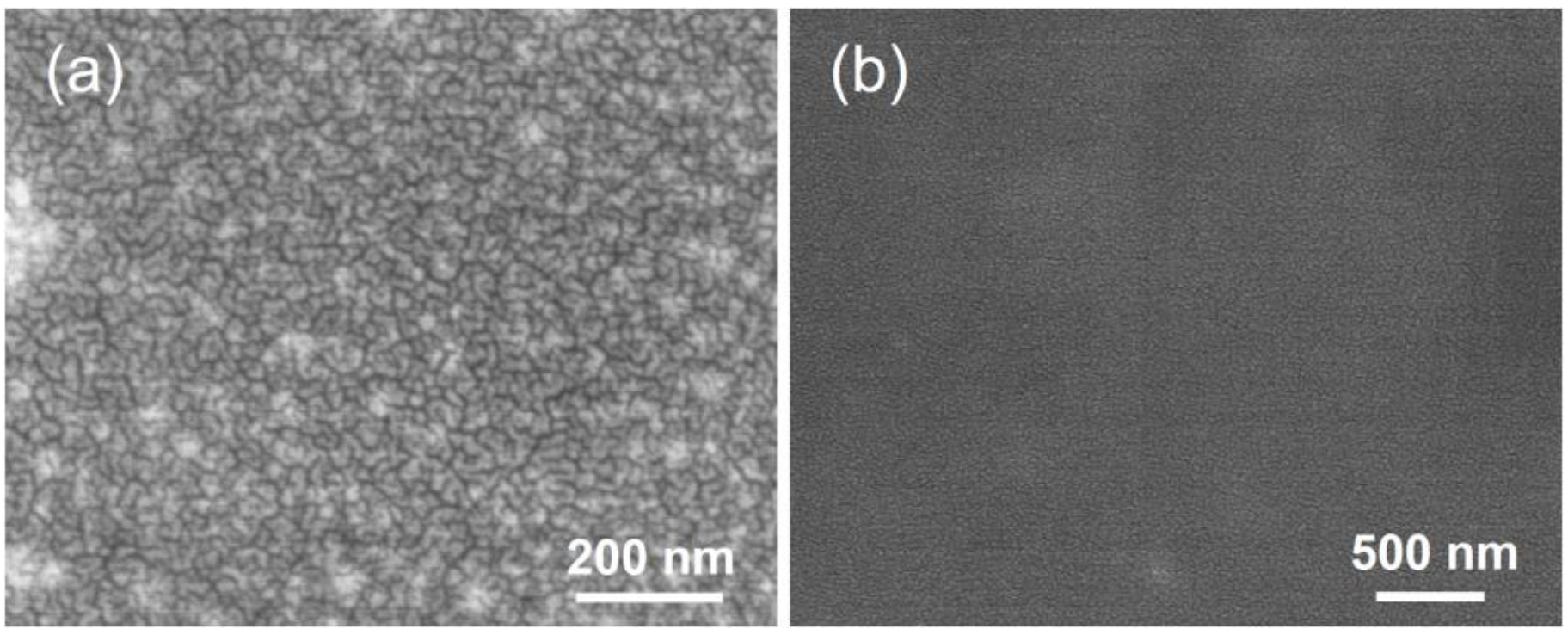

**Figure S2.** Surface SEM images of the PS-*b*-P2VP-UPCFs/AAO composite membranes. (a) 0.03 wt%. (b) 0.1 wt%.

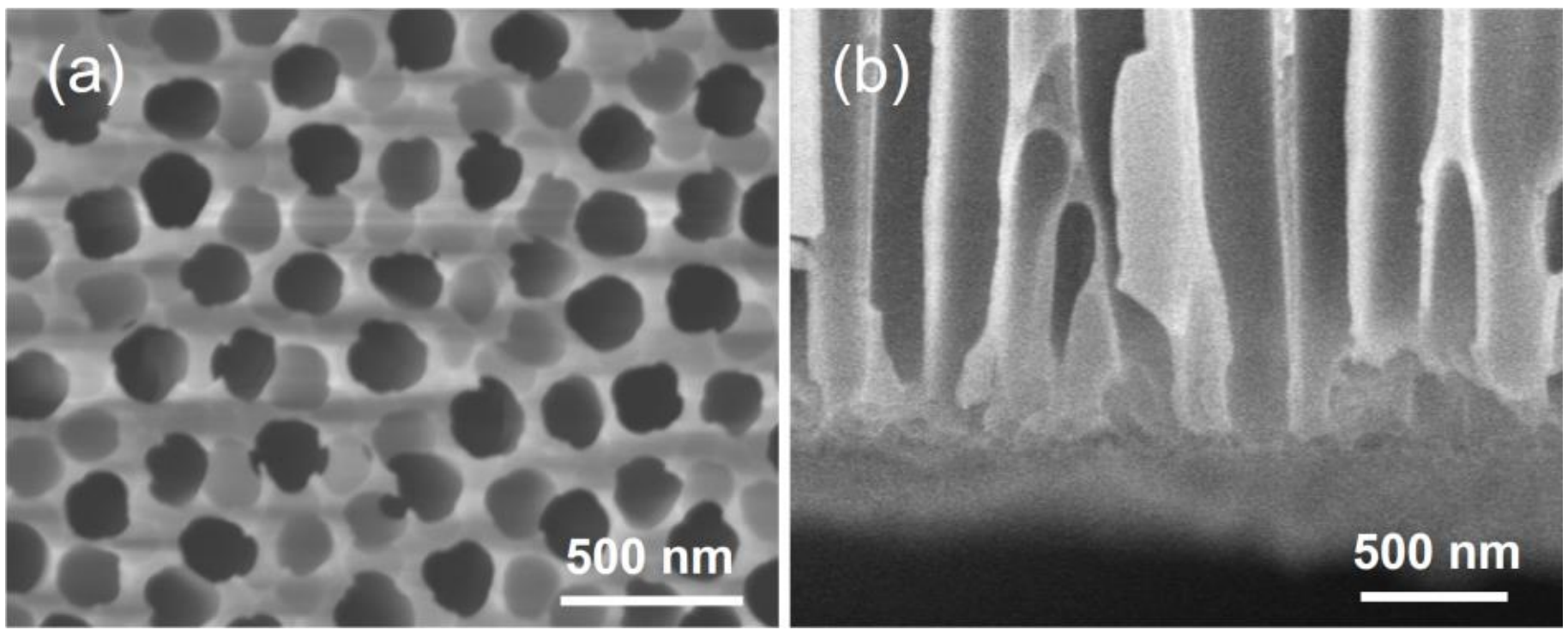

**Figure S3.** SEM images of the PS-*b*-P2VP-UPCFs/AAO composite membrane prepared with a concentration of 0.05 wt%. (a) Bottom surface. (b) Bottom cross section.

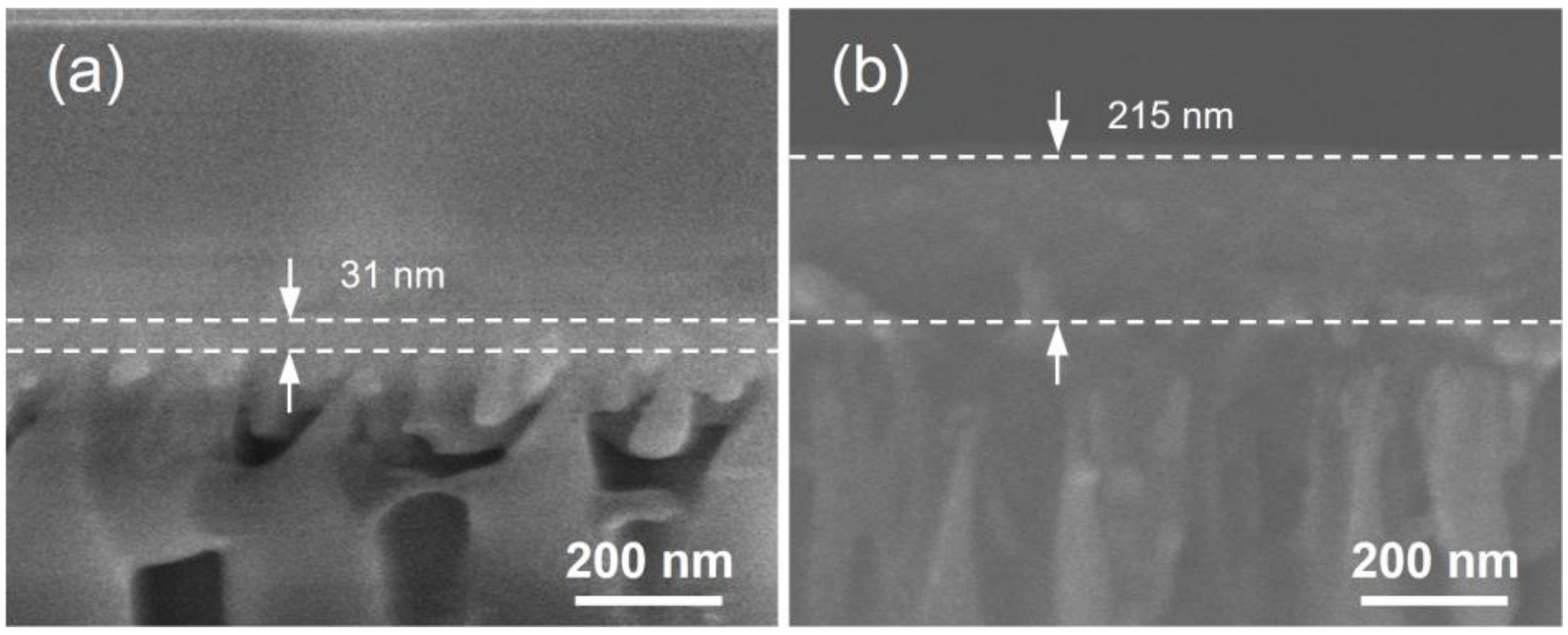

**Figure S4.** Cross-sectional SEM images of the PS-*b*-P2VP-UPCFs/AAO composite membranes. (a) 0.03 wt%. (b) 0.1 wt%.

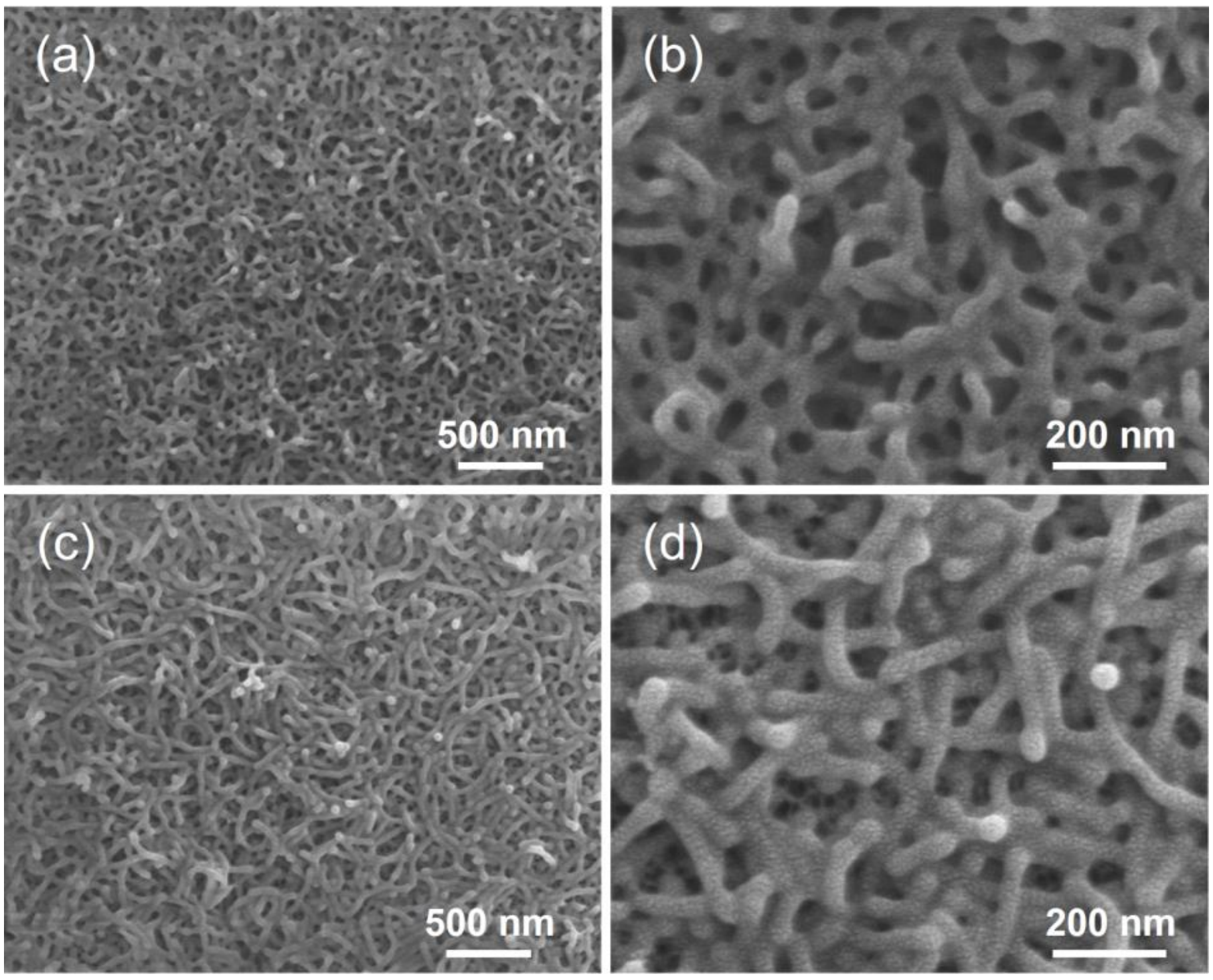

**Figure S5.** Surface SEM images of the PS-*b*-P2VP-*m*UPCFs/AAO composite membranes prepared with a concentration of 0.05 wt%. (a) 65°C for 1 h. (b) 75°C for 1 h.

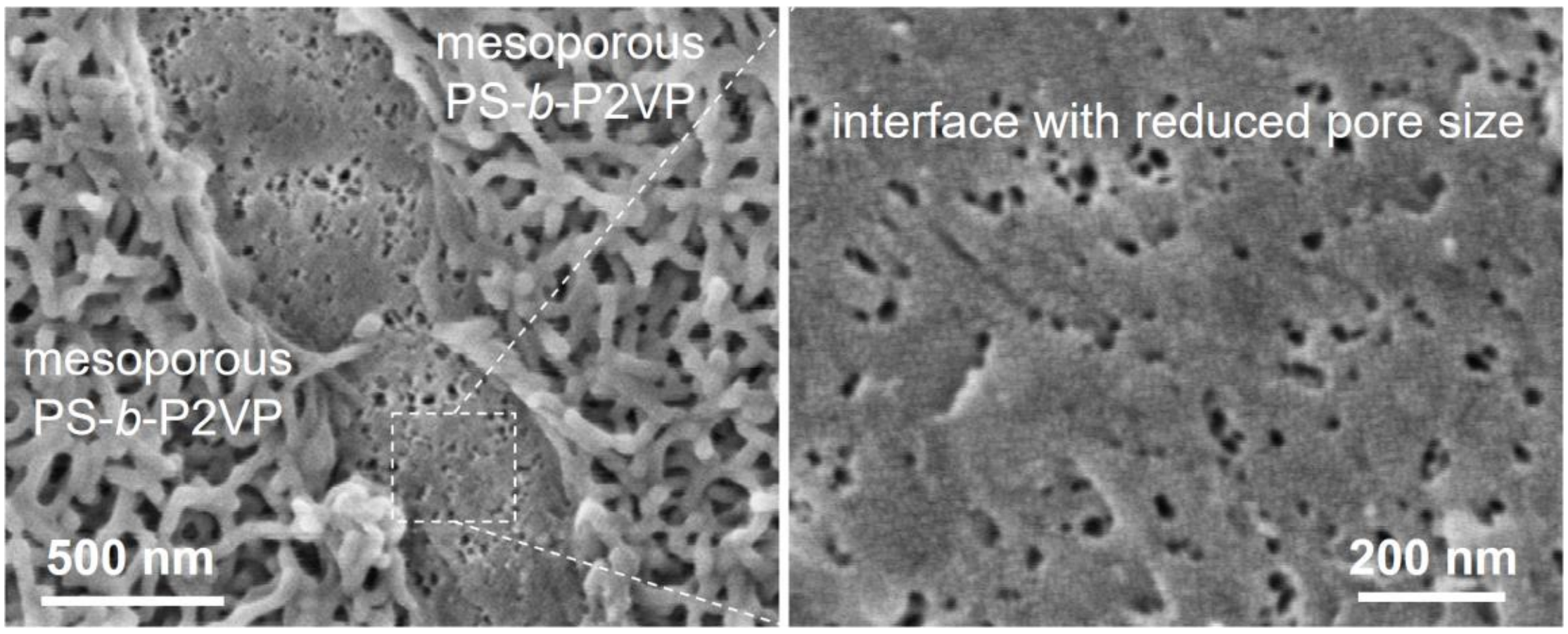

**Figure S6.** Surface SEM images of the PS-*b*-P2VP-*m*UPCFs/AAO composite membrane after removing the top mesoporous PS-*b*-P2VP layer. The membrane was prepared with concentration of 0.05 wt% followed by swelling at 70°C for 1 h.

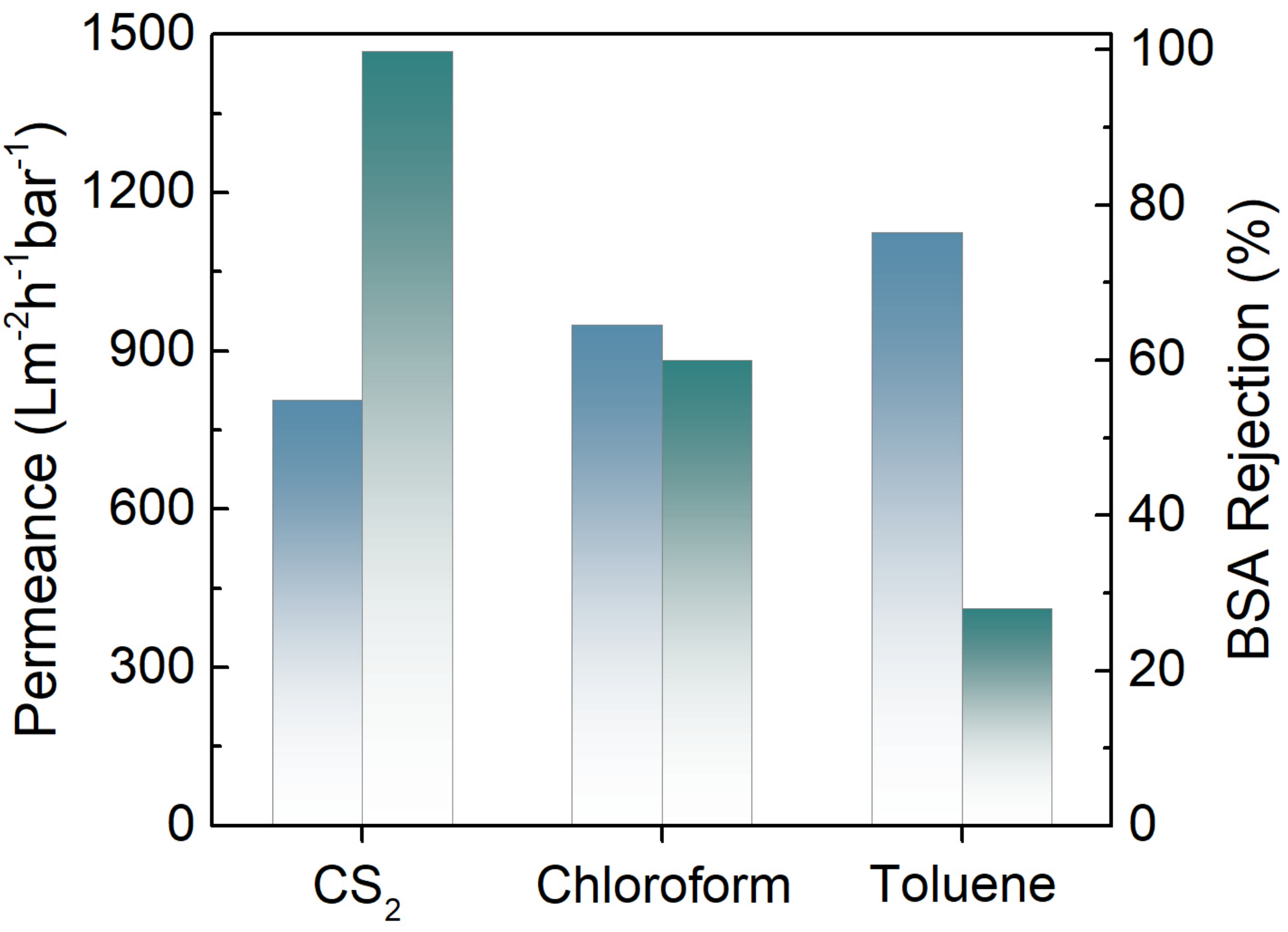

**Figure S7.** Separation performances of the PS-*b*-P2VP-*m*UPCFs/AAO composite membranes prepared with different solvent.